\documentclass[sigconf,authorversion]{acmart}

\usepackage{multirow}
\usepackage{enumitem}
\usepackage{makecell}

\usepackage{hyperref}
\usepackage{bm}
\usepackage{float}
\usepackage{subcaption}
\newcolumntype{?}{!{\vrule width 1.5pt}}

\definecolor{rulegray}{gray}{0.70}
\AtBeginDocument{%
  }

\setcopyright{acmlicensed}
\copyrightyear{2025}
\acmYear{2025}
\setcopyright{acmlicensed}\acmConference[RecSys '25]{Proceedings of the Nineteenth ACM Conference on Recommender Systems}{September 22--26, 2025}{Prague, Czech Republic}
\acmBooktitle{Proceedings of the Nineteenth ACM Conference on Recommender Systems (RecSys '25), September 22--26, 2025, Prague, Czech Republic}
\acmDOI{10.1145/3705328.3748038}
\acmISBN{979-8-4007-1364-4/2025/09}

\begin{document}

\title{Let It Go? Not Quite: Addressing Item Cold Start in Sequential Recommendations with Content-Based Initialization}

\author{Anton Pembek}
\email{apembek@bk.ru}
\affiliation{%
  \institution{Sber AI Lab, Lomonosov Moscow State University (MSU)}
  \city{Moscow}
  \country{Russian Federation}
}

\author{Artem Fatkulin}
\email{artem42fatkulin@gmail.com}
\affiliation{%
  \institution{Sber AI Lab, HSE University}
  \city{Moscow}
  \country{Russian Federation}
}

\author{Anton Klenitskiy}
\email{antklen@gmail.com}
\orcid{0009-0005-8961-6921}
\affiliation{
  \institution{Sber AI Lab}
  \city{Moscow}
  \country{Russian Federation}
}

\author{Alexey Vasilev}
\email{alexxl.vasilev@yandex.ru}
\orcid{0009-0007-1415-2004}
\affiliation{
  \institution{Sber AI Lab, HSE University}
  \city{Moscow}
  \country{Russian Federation}
}

\begin{abstract}

Many sequential recommender systems suffer from the cold start problem, where items with few or no interactions cannot be effectively used by the model due to the absence of a trained embedding. Content-based approaches, which leverage item metadata, are commonly used in such scenarios. One possible way is to use embeddings derived from content features such as textual descriptions as initialization for the model embeddings. However, directly using frozen content embeddings often results in suboptimal performance, as they may not fully adapt to the recommendation task. On the other hand, fine-tuning these embeddings can degrade performance for cold-start items, as item representations may drift far from their original structure after training.

We propose a novel approach to address this limitation. Instead of entirely freezing the content embeddings or fine-tuning them extensively, we introduce a small trainable delta to frozen embeddings that enables the model to adapt item representations without letting them go too far from their original semantic structure. This approach demonstrates consistent improvements across multiple datasets and modalities, including e-commerce datasets with textual descriptions and a music dataset with audio-based representation.

\end{abstract}

\begin{CCSXML}
<ccs2012>
  <concept>
   <concept_id>10002951.10003317.10003347.10003350</concept_id>
   <concept_desc>Information systems~Recommender systems</concept_desc>
  <concept_significance>500</concept_significance>
 </concept>
</ccs2012>
\end{CCSXML}

\ccsdesc[500]{Information systems~Recommender systems}

\keywords{Recommender Systems, Sequential Recommendations, Cold-start}

\begin{teaserfigure}
    \includegraphics[width=\textwidth]{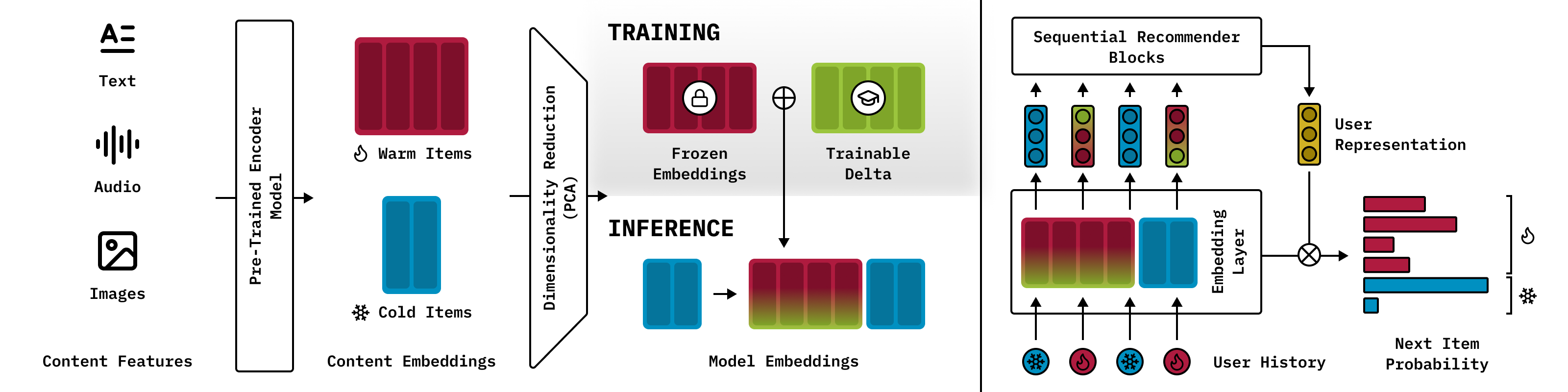}
    \caption{Illustration of the proposed approach.}
    \Description{Proposed approach.}
    \label{fig:schema}
\end{teaserfigure}
\maketitle

\section{Introduction}
The cold start problem remains a significant challenge for recommender systems in general and sequential recommender systems in particular. These systems aim to predict a user's next interaction based on their history, but struggle when encountering cold items with few or no interactions. This lack of interactions means the model cannot learn effective representations for these items, leading to poor recommendation quality.

A common strategy to address this issue is to leverage content-based features, such as textual descriptions, to construct embeddings for cold items. However, this approach introduces a distribution gap: the embeddings of warm items, learned during model training, may differ significantly from content-based embeddings used for cold items. To bridge this gap, various techniques have been explored, including learning transformations from cold embeddings to the warm item embedding space~\cite{van2013deep,volkovs2017dropoutnet,zhu2021learning}, employing contrastive learning~\cite{wei2021contrastive}, adversarial training~\cite{chen2022generative}, and distillation~\cite{huang2023aligning} to encourage similarity between cold and warm item representations.

Recent works~\cite{harte2023leveraging,boz2024improving} demonstrated that using text embeddings as initialization for transformer-based sequential recommendation models like SASRec~\cite{kang2018self} and BERT4Rec~\cite{sun2019bert4rec} improves overall recommendation quality. In our work, we investigate how content-based initialization, not limited to text but also including audio-based features, affects the cold start problem in sequential recommendation. A straightforward solution is to substitute content embeddings for cold items into a trained model. However, allowing full fine-tuning of these embeddings during training can hurt performance on cold items, as the embeddings may drift significantly from their initial content-based representation. On the other hand, fully freezing content embeddings limits the model's flexibility and hurts overall performance, as it cannot fully adapt to the interaction data.

To address these limitations, we propose an approach where item embeddings consist of two components. The first component is frozen content embeddings with a fixed norm. The second component is a small trainable delta vector with a bounded norm. This setup allows the model to adapt item representations to the interaction data without letting them go far from their original semantic structure derived from content. As a result, it improves recommendation quality on new items without sacrificing performance on those seen during training.

The main contributions of our work are as follows:
\begin{itemize}
    \item We investigate the impact of content-based embedding initialization on the cold start problem in transformer-based sequential recommendation.
    \item We propose a method that learns a small trainable delta with bounded norm on top of frozen content embeddings.
    \item We demonstrate that this approach consistently improves performance on cold items across different data modalities, including textual item descriptions and audio representations of songs.
\end{itemize}

\section{Related Work}

Using content-based representations to derive model embeddings is a common strategy for addressing the item cold start problem. Different approaches have been proposed to align cold item embeddings with the warm embedding space.
The work~\cite{van2013deep} predicts latent factors directly from a song's audio content. DropoutNet~\cite{volkovs2017dropoutnet} takes both latent preference factors and content features as input and applies input dropout to the latent factors during training, forcing the model to rely on content when preference information is missing. The paper~\cite{zhu2021learning} uses meta networks to generate item-specific scaling and shifting functions that transform cold item embeddings into the warm feature space. CLCRec~\cite{wei2021contrastive} applies contrastive learning to maximize the mutual information between item content representations and collaborative embeddings. GAR~\cite{chen2022generative} employs adversarial training between a generator and a recommender to ensure that generated cold item embeddings have a similar distribution to warm ones. ALDI~\cite{huang2023aligning} introduces a distillation framework where warm items act as "teachers" and cold items as "students," aligning the students' content-based predictions with the teachers' behavior-based predictions. These works mainly focus on collaborative filtering models rather than sequential recommendations.

For sequential recommendations, M2TRec~\cite{shalaby2022m2trec} introduces an item-ID-free framework that learns item representations directly from metadata and uses multi-task learning. Recformer~\cite{li2023text} models items and user preferences using language representations derived solely from item textual attributes. This approach, however, increases computational complexity compared to ID-based methods due to much longer input sequences. In contrast to these works, we focus on leveraging content embeddings within classic ID-based models like SASRec. SimRec~\cite{brody2024simrec} and the work~\cite{wang2024language} incorporate item similarities derived from textual embeddings into the training process of such models using customized loss functions. Unlike our approach, they consider tail items - those with very few interactions but not completely new - as cold. The works~\cite{harte2023leveraging,boz2024improving} use textual embeddings to initialize the transformer embedding layer and show that this approach improves recommendation metrics. However, they do not specifically address the cold start problem. The paper~\cite{tamm2024comparative} explores the usage of frozen pretrained audio representations for music recommendations without any fine-tuning.

\section{Approach}
\label{sec:approach}
\subsection{Task formulation}

Let $\mathcal{U}$ be the set of users and $\mathcal{I}$ be the set of items. In the sequential recommendation setting, we are given an ordered history of user interactions with items. The objective is to predict the next item the user will interact with. State-of-the-art sequential recommender systems (e.g., SASRec and BERT4Rec) typically define a learnable embedding function $E: \mathcal{I} \to \mathbb{R}^m$, which maps each item $i \in \mathcal{I}$ to a vector representation $\mathbf{e}^i \in \mathbb{R}^m$. A transformer-based architecture aggregates a sequence of user interactions into a user representation $\mathbf{h}^u \in \mathbb{R}^m$. This representation is then used in a Maximum Inner Product Search (MIPS) to compute relevance scores for all items:
\begin{equation}
    r(u, i) = \mathbf{h}^u \cdot \mathbf{e}^i
\label{eq:mips}
\end{equation}

While demonstrating superior overall performance, such models fail to handle new items due to the absence of trained embeddings for them — a problem called item cold start. At the same time, enabling the model to make use of cold items can increase recommendation diversity, improve overall performance by leveraging interactions with newly introduced content, and extend the model's deployment lifetime by reducing the need for frequent retraining. Addressing this limitation is therefore crucial for building robust and efficient recommender systems.

\subsection{Content-based initialization}

In many domains, items are accompanied by additional information, such as textual descriptions for goods or sound-based embeddings for music tracks. In the absence of trained embeddings for new items, it is reasonable to utilize this content information to generate initial representations. For cold items, these representations can be used as-is during inference, while for warm items, they can be fine-tuned along with other parts of the model during training. Prior works have shown~\cite{harte2023leveraging,boz2024improving} that initializing the model with text-based embeddings can significantly improve recommendation quality. Nevertheless, the effectiveness of content-based initialization depends critically on both the quality of the source information and the capabilities of the encoder model. For example, as previously shown in~\cite{tamm2024comparative}, different audio encoding techniques demonstrate various performance when used to generate music track representations.

Furthermore, the direct use of content-based embeddings comes with a major drawback: freezing them during training leads to degraded model performance, while allowing them to be updated makes the approach less suitable for the item cold start scenario. That is, once training is complete, the content-based embeddings of cold items remain outside the learned representation space, resulting in poor generalization to unseen items.

\subsection{Representation adjustment}

To overcome this limitation, we propose freezing content-based item embeddings, fixing their norm, and training only a small delta layer. This allows the model to adjust item representations slightly while keeping them close to the original embeddings.

\begin{figure}[H]
\includegraphics{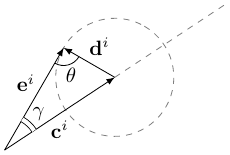}
\caption{Geometric representation of the proposed approach. $\mathbf{c}^i$ is a frozen content-based item embedding, $\mathbf{d}^i$ is the corresponding trainable correction vector and $\mathbf{e}^i$ is the final item representation.}
\label{fig:adjusted_representation}
\Description{Geometric representation of the proposed approach.}
\end{figure}

Formally, given a content-based embedding $\mathbf{c}^i \in \mathbb{R}^m$ of an item $i \in \mathcal{I}$ with unit norm $\|\mathbf{c}^i\| = 1$, and a trainable delta vector $\mathbf{d}^i \in \mathbb{R}^m$ with $\|\mathbf{d}^i\| = \delta_i$, where $0 \leq \delta_i < 1$, we consider the cosine similarity between the original embedding and its adjusted form $\mathbf{e}^i = \mathbf{c}^i + \mathbf{d}^i$:
\begin{equation}
    \mathrm{sim}(\mathbf{c}^i, \mathbf{e}^i) =
    \cos\gamma = \sqrt{1 - \sin^2\gamma}
    \label{eq:cosine_similarity}
\end{equation}
where $\gamma$ denotes the angle between $\mathbf{c}^i$ and $\mathbf{e}^i$. 

Although the minimum cosine similarity can be obtained by straightforward differentiation, here we present a more elegant and intuitive derivation using the law of sines:
\begin{equation}
\frac{\|\mathbf{d}^i\|}{\sin\gamma} =
\frac{\|\mathbf{c}^i\|}{\sin\theta}
\quad\Longrightarrow\quad
\sin\gamma = \delta_i\,\sin\theta
\label{eq:law_of_sines}
\end{equation}
where $\theta$ is the interior angle of the triangle opposite the side represented by vector $\mathbf{c}^i$ (see Figure~\ref{fig:adjusted_representation}).

Applying this substitution to Equation~\ref{eq:cosine_similarity} gives an explicit relationship between cosine similarity and the norm of the correction vector:
\begin{equation}
    \mathrm{sim}(\mathbf{c}^i, \mathbf{e}^i) =
    \sqrt{1 - \delta_i^2 \sin^2\theta}
\end{equation}
The minimum cosine similarity occurs when $\theta = \pi / 2$ and is equal: 
\begin{equation}
    \min_{\theta} \mathrm{sim}(\mathbf{c}^i, \mathbf{e}^i) = \sqrt{1 - \delta_i^2}
\label{eq:min_cosine_similarity}
\end{equation}

Geometrically, if we visualize the endpoint of $\mathbf{e}^i$ tracing a sphere of radius $\delta_i$ around the endpoint of $\mathbf{c}^i$, the case where the vector $\mathbf{d}^i$ is orthogonal to $\mathbf{e}^i$ corresponds to this minimum (see Figure~\ref{fig:adjusted_representation}).

Thus, by varying the norm $\delta_i$, we can control how close the item representation is to the original content-based embedding. We apply the following strategy to constrain the correction vectors during training: we introduce a hyperparameter $\delta_{\text{max}}$ and clip the norm of each vector if it exceeds this threshold.

In the item cold start scenario, maintaining proximity to content-based embeddings enables direct use of these representations for cold items.
Additionally, the proposed technique also serves as a form of regularization. Since sequential recommenders rely on MIPS, where the embedding norm directly influences item scores, it is important to manage the variation in norms across items, which we found to be substantial.

\section{Experiments}

\subsection{Experimental settings}

\subsubsection{Datasets}

\begin{table}[!htbp]
\setlength{\abovecaptionskip}{2pt}
\setlength{\belowcaptionskip}{-7pt}
\caption{\textbf{Statistics of the datasets after preprocessing, including average sequence length and the percentage of cold items in the ground truth (GT).}}
\resizebox{\columnwidth}{!}{%
\label{tab:datasetStats}
    \centering
    \begin{tabular}{lrrrrr}
    \hline 
        \multirow{2}{*}{\textbf{Dataset}} & \multirow{2}{*}{\textbf{\# Users}} & \multirow{2}{*}{\textbf{\# Items}} & \multirow{2}{*}{\textbf{\# Interact.}} & \multicolumn{1}{c}{\textbf{Avg.}} & \multicolumn{1}{c}{\textbf{\ Cold items}} \\ 
        & & & & \multicolumn{1}{c}{\textbf{length}} & \multicolumn{1}{c}{\textbf{ in GT}} \\ \hline 
        \textbf{Amazon-M2 FR \cite{jin2023amazon}} &  129,983 & 44,049 & 566,806 & 4.3 & 7\% \\
        
        \textbf{Beauty \cite{mcauley2015image}}  & 21,029 & 11,733 & 149,147 & 7.1 & 25\% \\ 
        
        \textbf{Zvuk \cite{shevchenko2024variability}} & 9,076 & 131,085 & 3,236,653  & 356.6 & 13\% \\
        
    \hline 
    \end{tabular}
    }
\end{table}

\renewcommand{\arraystretch}{1.8}
\setlength{\tabcolsep}{4pt}

\begin{table*}[htbp]
  \caption{
    \textbf{Performance on cold, warm, and all ground‑truth items.  
    Bold numbers mark the best model; the second best is \underline{underlined}.
    Abbreviations: c.i. -- content initialization, t.d. -- trainable delta, GT -- ground truth.}
  }
  \label{tab:results}
  \centering
  \resizebox{\textwidth}{!}{
      \begin{tabular}{ll*{9}{c}}
        \toprule
        \multirow{2}{*}{\textbf{Metric}} & \multicolumn{1}{c}{\multirow{2}{*}{\textbf{Model}}} &
        \multicolumn{3}{c}{\textbf{Amazon‑M2}} &
        \multicolumn{3}{c}{\textbf{Beauty}} &
        \multicolumn{3}{c}{\textbf{Zvuk}} \\ 
        \cmidrule{3-11}
        ~ & ~ &
        \textbf{Cold GT} & \textbf{Warm GT} & \textbf{Total} &
        \textbf{Cold GT} & \textbf{Warm GT} & \textbf{Total} &
        \textbf{Cold GT} & \textbf{Warm GT} & \textbf{Total} \\
        \midrule
        \multirow{4}{*}{\textbf{HR@10}}
          & Content‑based KNN                 & \underline{0.454$\pm$0.000} & 0.383$\pm$0.000 & 0.388$\pm$0.000 & \textbf{0.043$\pm$0.000} & 0.044$\pm$0.000 & 0.044$\pm$0.000 & 0.009$\pm$0.000 & 0.000$\pm$0.000 & 0.001$\pm$0.000 \\
          & SASRec                            & 0.000$\pm$0.000 & 0.610$\pm$0.003 & 0.567$\pm$0.002 & 0.000$\pm$0.000 & 0.072$\pm$0.001 & 0.054$\pm$0.001 & 0.000$\pm$0.000 & \textbf{0.094$\pm$0.003} & \underline{0.082$\pm$0.003} \\
          & SASRec with c.i.             & 0.435$\pm$0.015 & \textbf{0.620$\pm$0.002} & \underline{0.607$\pm$0.001} & 0.032$\pm$0.004 & \underline{0.088$\pm$0.002} & \underline{0.074$\pm$0.002} & \underline{0.023$\pm$0.003} & 0.088$\pm$0.002 & 0.080$\pm$0.002 \\
          & \textit{SASRec with t.d. (ours)} & \textbf{0.509$\pm$0.005} & \underline{0.617$\pm$0.002} & \textbf{0.609$\pm$0.002} & \underline{0.038$\pm$0.008} & \textbf{0.092$\pm$0.002} & \textbf{0.078$\pm$0.003} & \textbf{0.034$\pm$0.002} & \textbf{0.094$\pm$0.002} & \textbf{0.087$\pm$0.002} \\
        \midrule
        \multirow{4}{*}{\textbf{NDCG@10}}
          & Content‑based KNN                 & \underline{0.312$\pm$0.000} & 0.232$\pm$0.000 & 0.238$\pm$0.000 & \textbf{0.022$\pm$0.000} & 0.024$\pm$0.000 & 0.024$\pm$0.000 & 0.006$\pm$0.000 & 0.000$\pm$0.000 & 0.001$\pm$0.000 \\
          & SASRec                            & 0.000$\pm$0.000 & 0.438$\pm$0.002 & 0.407$\pm$0.002 & 0.000$\pm$0.000 & 0.043$\pm$0.001 & 0.033$\pm$0.001 & 0.000$\pm$0.000 & \textbf{0.063$\pm$0.001} & \textbf{0.055$\pm$0.001} \\
          & SASRec with c.i.             & 0.297$\pm$0.010 & \textbf{0.448$\pm$0.001} & \underline{0.438$\pm$0.000} & 0.018$\pm$0.002 & \underline{0.053$\pm$0.001} & \underline{0.044$\pm$0.001} & \underline{0.014$\pm$0.002} & 0.058$\pm$0.002 & 0.052$\pm$0.001 \\
          & \textit{SASRec with t.d. (ours)} & \textbf{0.359$\pm$0.000} & \underline{0.446$\pm$0.002} & \textbf{0.440$\pm$0.002} & \textbf{0.022$\pm$0.004} & \textbf{0.054$\pm$0.001} & \textbf{0.046$\pm$0.002} & \textbf{0.021$\pm$0.002} & \underline{0.060$\pm$0.002} & \textbf{0.055$\pm$0.002} \\
        \bottomrule
      \end{tabular}
    }
\end{table*}

We evaluate our approach on three datasets from different domains. Amazon-M2~\cite{jin2023amazon} is a dataset from the KDD Cup 2023 competition\footnote{\url{https://kddcup23.github.io}}. It comprises customer shopping sessions from six locales and includes textual item descriptions. Beauty, one of the most widely used datasets in sequential recommendation with inherent sequential patterns~\cite{klenitskiy2024does}, contains customer reviews and is also rich in textual information. To verify that our method generalizes across modalities, we additionally examine Zvuk~\cite{shevchenko2024variability}, a dataset with strong sequential structure~\cite{klenitskiy2024does} from the music streaming service containing audio-based item representations. 

For Amazon-M2, we use the original data provided by the authors, restricting our evaluation to the France locale (specifically, Task 2, Phase 1). Following~\cite{hidasi2023widespread,klenitskiy2024does}, we remove consecutive duplicated items from the user sequence for all datasets.
For the Zvuk dataset, we consider interactions longer than 60 seconds as positive and retain only those, while for Beauty, we filter out reviews with a rating lower than 4. Additionally, for Zvuk, we randomly sample 10,000 users to obtain a representative yet computationally efficient subset. Finally, we apply $n$-core filtering \cite{sun2020we} with $n=3$  to the train/validation data for Zvuk and Beauty. The final statistics of the datasets after preprocessing are summarized in Table~\ref{tab:datasetStats}.

\subsubsection{Evaluation}

For Amazon-M2, we use the original train-test split configuration with a 2-week training period followed by a 1-week test period. For Beauty and Zvuk, we use a global temporal split to prevent data leakage~\cite{ji2023critical,sun2023take} with the temporal boundary set at 90\% of the interactions. The validation set consists of sequences from 10\% of randomly sampled users for all datasets. The last interaction is used as the ground truth for validation and test sets, while the previous part of the sequence is used as input to the model. This setup is appropriate for next-item prediction according to~\cite{seqsplits2025}.

To evaluate the quality of the recommendation, we use Normalized Discounted Cumulative Gain (NDCG@10) and Hit Rate (HR@10) from the Replay library \cite{vasilev2024replay}. We distinguish between warm items, which are present in the training set, and cold items appearing only in the test set. To evaluate the impact on the cold-start problem, we measure the performance separately for users with warm ground-truth items and users with cold ground-truth items. Table~\ref{tab:datasetStats} shows the proportion of such users.

\subsubsection{Implementation Details}
We perform the experiments with the SASRec model~\cite{kang2018self}. All models are trained using full cross-entropy loss following~\cite{klenitskiy2023turning}, with a batch size of 128. We employ the Adam optimizer with a learning rate of 1e-3. SASRec is configured with two transformer blocks, a single attention head, and a dropout rate of 0.3. The embedding dimension is set to 128 for Beauty and Zvuk, and to 64 for Amazon-M2. The maximum input sequence length is limited to 128 for Zvuk, and to 64 for Amazon-M2 and Beauty. For text-based content embeddings, we employ the E5 encoder \cite{wang2022text}, while the Zvuk dataset includes precomputed audio embeddings \footnote{\url{https://www.kaggle.com/datasets/alexxl/zvuk-dataset}}.

In practice, content-based embeddings often exhibit higher dimensionality than the model's internal representations. To reconcile this mismatch, we first apply component-wise standardization, followed by Principal Component Analysis (PCA) to reduce the embeddings to the target dimension.

To increase the statistical significance of the results, we run each experiment five times with different random seeds and calculate aggregated metrics. All code necessary to reproduce our experiments is available at our GitHub repository\footnote{\url{https://github.com/ArtemF42/let-it-go}}.

\subsection{Results}

\subsubsection{Main results}
\label{sec:main_results}

Table \ref{tab:results} summarizes the results of the experiments on all datasets. The \textit{Content-based KNN} method serves as a content-based baseline that recommends items most similar to the average content embedding of the user's sequence. \textit{SASRec} refers to the standard version of the model, which cannot handle cold items. \textit{SASRec with content initialization} is a baseline variant where item embeddings are initialized from content and then fully fine-tuned during training. \textit{SASRec with trainable delta} is our proposed approach from Section~\ref{sec:approach} with a maximum norm constraint of the delta vector $\delta_{\text{max}} = 0.5$ corresponding to minimum cosine similarity of $0.87$.

\textit{SASRec with content initialization} outperforms the content-based baseline on the Zvuk dataset and performs slightly worse on the other datasets in the cold-item settings. As expected, the content-based baseline performs significantly worse on warm items. Importantly, incorporating cold items into SASRec does not degrade its performance on warm items. Content initialization even improves warm-item metrics compared to the original model on Amazon-M2 and Beauty datasets.

Our proposed version with a trainable delta leads to substantial improvements in cold-item metrics across all three datasets, outperforming the content-based baseline. Warm-item and overall performance also show slight improvements, demonstrating the robustness of our method.

\subsubsection{Norm of the delta vector}
\begin{figure}[ht!]
    \centering
 \begin{subfigure}{\columnwidth} 

    \includegraphics[width=1\linewidth]{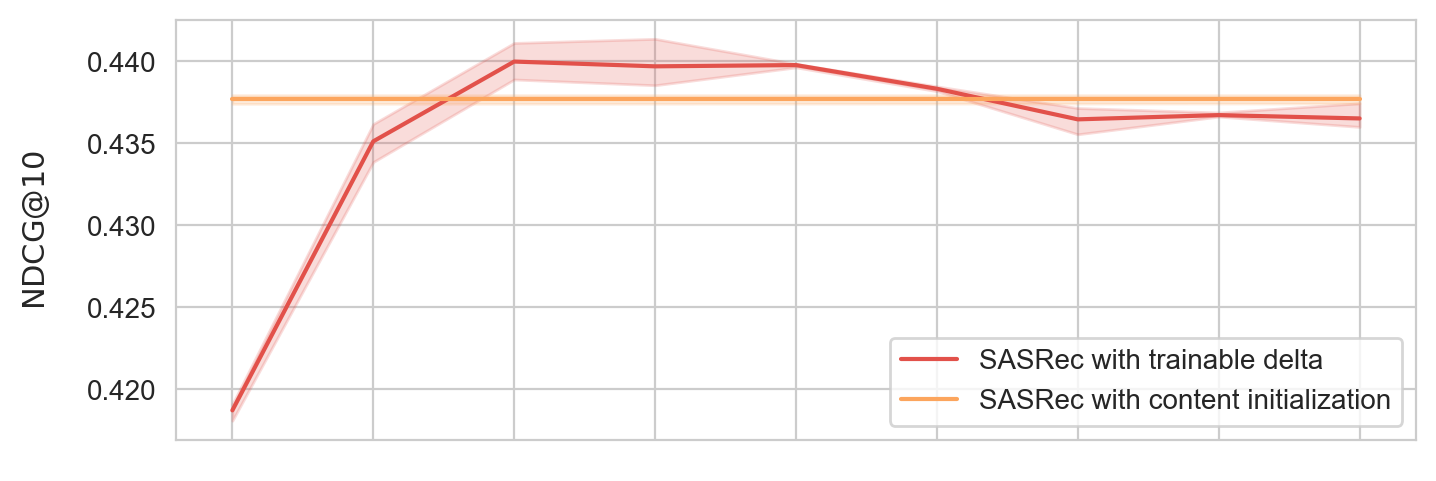}
 \end{subfigure}
 \begin{subfigure}{\columnwidth} 
    \includegraphics[width=1\linewidth]{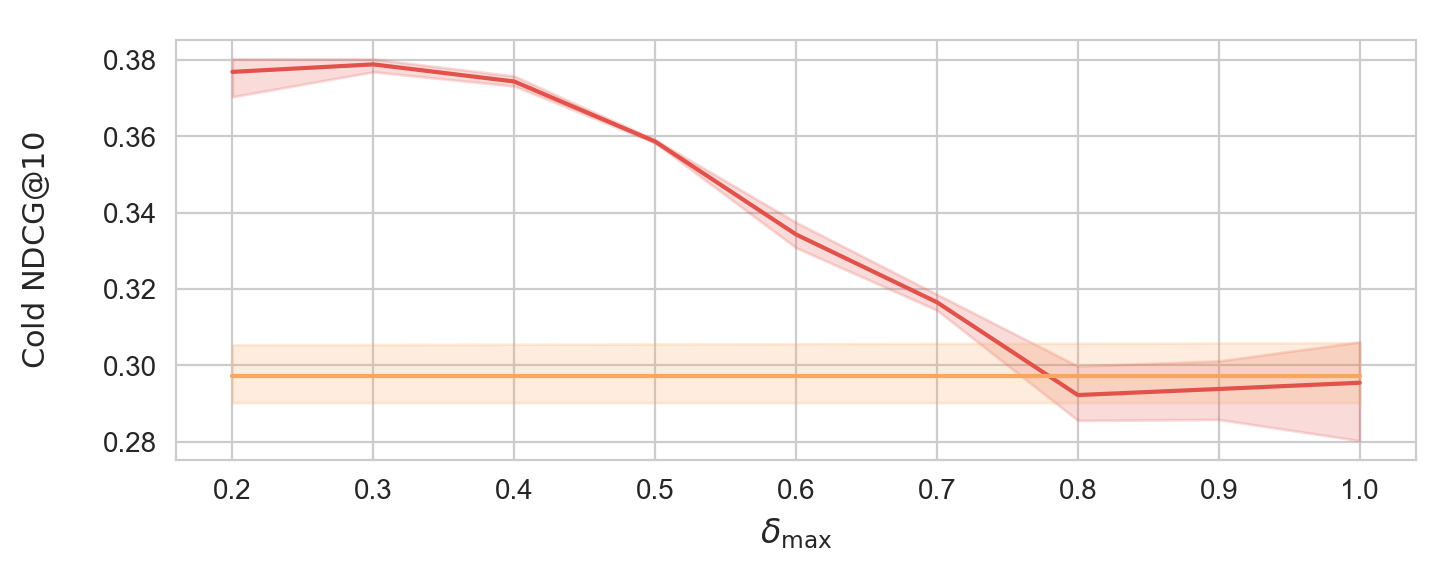}
    
   \end{subfigure}
    \caption{
        Mean total (top) and cold (bottom) NDCG@10 for SASRec with trainable delta evaluated against $\delta_{\text{max}}$ on the Amazon‑M2 dataset. SASRec with content initialization is provided for comparison. Low $\delta_{\text{max}}$ values yield high cold‑item metrics but degrade the model’s overall quality.
    }
    \Description{NDCG@10 comparison across warm/cold item ratios}

     \label{fig:norm_dependency}
    
\end{figure}

The norm of the trainable delta vector is the main hyperparameter in our approach. Figure~\ref{fig:norm_dependency} shows how NDCG@10 for cold and all items changes with different values of $\delta_{\text{max}}$ on the Amazon-M2 dataset. When $\delta_{\text{max}}$ is too small, the model does not have enough flexibility to adjust item representations, and the total quality drops significantly. When the value is too large, the embeddings move too far from their original initialization, which leads to worse performance on cold items. We find that values in the range of 0.3 to 0.6 provide a good trade-off, giving the model enough capacity to learn while still keeping the embeddings close to their initialization.

\subsubsection{Cold items in input sequences}

\begin{figure}[ht!]
    \resizebox{1\columnwidth}{!}{
    \centering
    \includegraphics[width=1\linewidth]{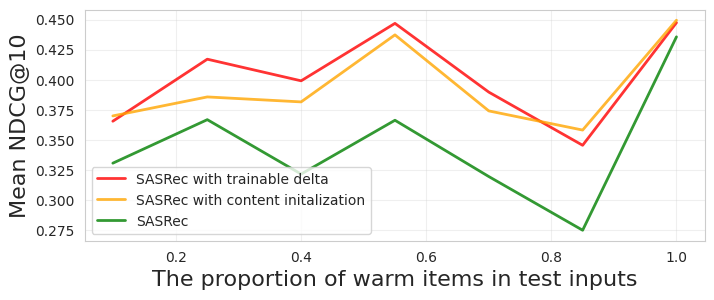}
    }
    \caption{
        Mean NDCG@10 for SASRec, SASRec with content initialization, and SASRec with trainable delta, evaluated across different proportions of warm items in test input sequences on Amazon-M2 dataset.
    }
    \Description{Change in NDCG@10 across different $\delta_{\text{max}}$ values}

    \label{fig:cold_input_items}
\end{figure}

While the results in Table~\ref{tab:results} are focused on cold items in the ground truth, Figure~\ref{fig:cold_input_items} shows how performance metrics change with the proportion of cold items in user input sequences on the Amazon-M2 dataset. Using content-based initialization significantly improves recommendation quality when cold items are presented in the sequence. Implementing a trainable delta leads to a further, though smaller, improvement in this setting.

\subsubsection{Low-frequency items} 
\begin{figure}[t!]
    \centering
 \begin{subfigure}{\columnwidth} 

    \includegraphics[width=1\linewidth]{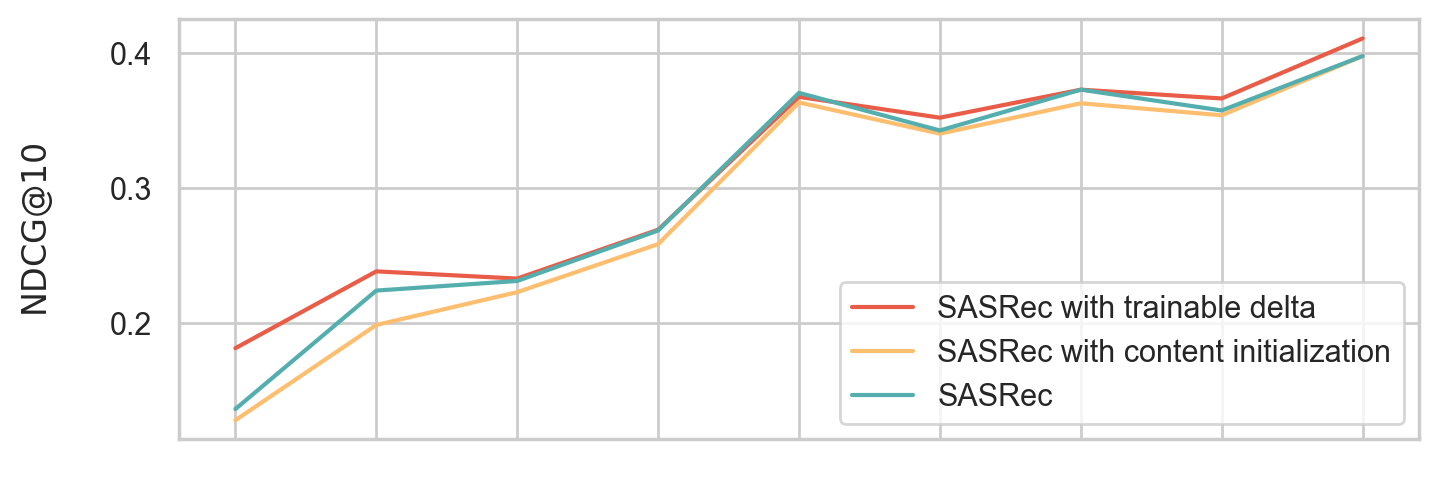}
 \end{subfigure}
 \begin{subfigure}{\columnwidth} 
    \includegraphics[width=1\linewidth]{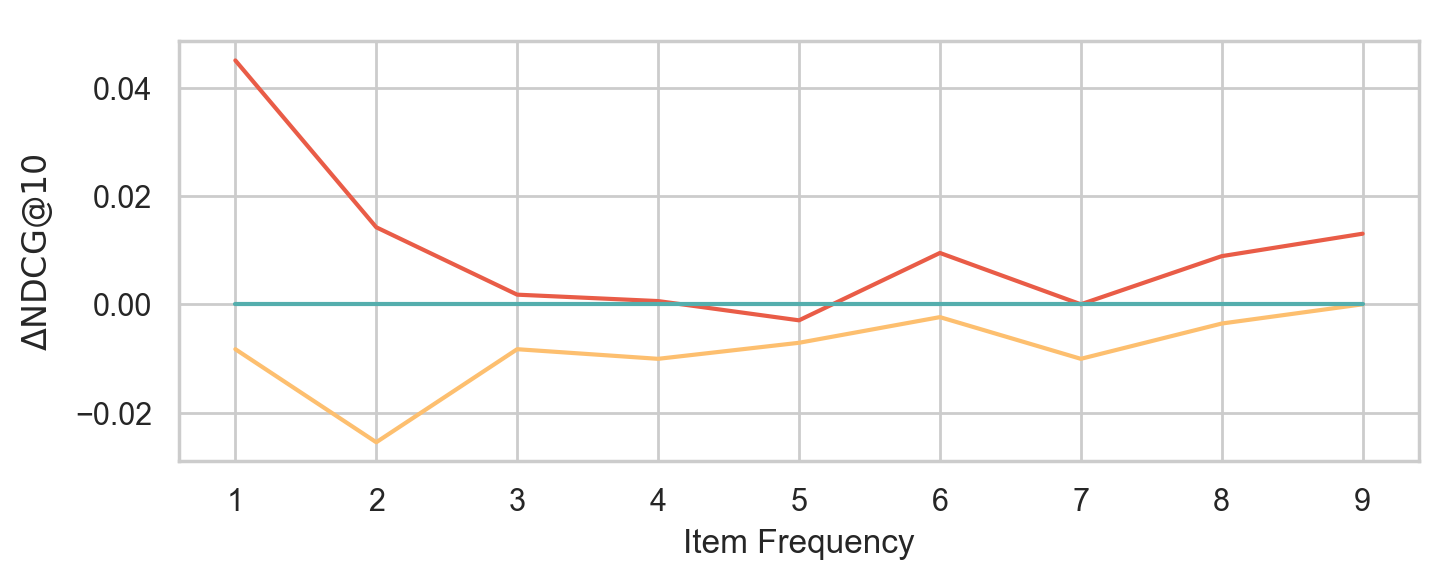}
    
   \end{subfigure}
    \caption{
        Mean NDCG@10 (top) and NDCG@10 relative to SASRec (bottom) for SASRec, SASRec with content initialization, and SASRec with trainable delta, evaluated against the frequency of ground-truth items in the training set on Amazon-M2 dataset.
    }
    \Description{Performance across item frequencies in training set}

    \label{fig:low_freq_items}
\end{figure}

We further analyze how recommendation quality metrics for ground-truth items vary with their frequency in the training set, as shown in Figure~\ref{fig:low_freq_items}. Our trainable delta approach demonstrates improvements for rare items, with performance gradually converging to SASRec levels as item frequency increases.

\section{Conclusion}

In this work, we proposed a simple yet effective method for addressing the cold start problem in sequential recommendations by adding a small trainable delta with a bounded norm to frozen content-based embeddings. We evaluated our approach on datasets with text-based and audio-based embeddings, confirming its applicability across different modalities of item content. Our key findings demonstrate that the performance on cold items in the ground truth shows significant improvement, while the metrics on warm items remain stable without degradation. 

Although the approach achieves superior quality metrics, it introduces additional training cost due to maintaining a second embedding vector for each item. This memory overhead may limit scalability for extremely large item sets. Future work could address this issue by reducing the embedding size or extending our study to additional modalities and recommendation scenarios.

\bibliographystyle{ACM-Reference-Format}
\bibliography{content/bibliography}

%%% -*-BibTeX-*-
%%% Do NOT edit. File created by BibTeX with style
%%% ACM-Reference-Format-Journals [18-Jan-2012].

\begin{thebibliography}{27}

%%% ====================================================================
%%% NOTE TO THE USER: you can override these defaults by providing
%%% customized versions of any of these macros before the \bibliography
%%% command.  Each of them MUST provide its own final punctuation,
%%% except for \shownote{} and \showURL{}.  The latter two
%%% do not use final punctuation, in order to avoid confusing it with
%%% the Web address.
%%%
%%% To suppress output of a particular field, define its macro to expand
%%% to an empty string, or better, \unskip, like this:
%%%
%%% \newcommand{\showURL}[1]{\unskip}   % LaTeX syntax
%%%
%%% \def \showURL #1{\unskip}           % plain TeX syntax
%%%
%%% ====================================================================

\ifx \showCODEN    \undefined \def \showCODEN     #1{\unskip}     \fi
\ifx \showISBNx    \undefined \def \showISBNx     #1{\unskip}     \fi
\ifx \showISBNxiii \undefined \def \showISBNxiii  #1{\unskip}     \fi
\ifx \showISSN     \undefined \def \showISSN      #1{\unskip}     \fi
\ifx \showLCCN     \undefined \def \showLCCN      #1{\unskip}     \fi
\ifx \shownote     \undefined \def \shownote      #1{#1}          \fi
\ifx \showarticletitle \undefined \def \showarticletitle #1{#1}   \fi
\ifx \showURL      \undefined \def \showURL       {\relax}        \fi
% The following commands are used for tagged output and should be
% invisible to TeX
\providecommand\bibfield[2]{#2}
\providecommand\bibinfo[2]{#2}
\providecommand\natexlab[1]{#1}
\providecommand\showeprint[2][]{arXiv:#2}

\bibitem[Boz et~al\mbox{.}(2025)]%
        {boz2024improving}
\bibfield{author}{\bibinfo{person}{Artun Boz}, \bibinfo{person}{Wouter Zorgdrager}, \bibinfo{person}{Zoe Kotti}, \bibinfo{person}{Jesse Harte}, \bibinfo{person}{Panos Louridas}, \bibinfo{person}{Vassilios Karakoidas}, \bibinfo{person}{Dietmar Jannach}, {and} \bibinfo{person}{Marios Fragkoulis}.} \bibinfo{year}{2025}\natexlab{}.
\newblock \showarticletitle{Improving sequential recommendations with llms}.
\newblock \bibinfo{journal}{\emph{ACM Transactions on Recommender Systems}} (\bibinfo{year}{2025}).
\newblock
\href{https://doi.org/10.1145/3711667}{doi:\nolinkurl{10.1145/3711667}}


\bibitem[Brody and Lagziel(2024)]%
        {brody2024simrec}
\bibfield{author}{\bibinfo{person}{Shaked Brody} {and} \bibinfo{person}{Shoval Lagziel}.} \bibinfo{year}{2024}\natexlab{}.
\newblock \showarticletitle{SimRec: Mitigating the cold-start problem in sequential recommendation by integrating item similarity}.
\newblock  (\bibinfo{year}{2024}).
\newblock
\urldef\tempurl%
\url{https://www.amazon.science/publications/simrec-mitigating-the-cold-start-problem-in-sequential-recommendation-by-integrating-item-similarity}
\showURL{%
\tempurl}


\bibitem[Chen et~al\mbox{.}(2022)]%
        {chen2022generative}
\bibfield{author}{\bibinfo{person}{Hao Chen}, \bibinfo{person}{Zefan Wang}, \bibinfo{person}{Feiran Huang}, \bibinfo{person}{Xiao Huang}, \bibinfo{person}{Yue Xu}, \bibinfo{person}{Yishi Lin}, \bibinfo{person}{Peng He}, {and} \bibinfo{person}{Zhoujun Li}.} \bibinfo{year}{2022}\natexlab{}.
\newblock \showarticletitle{Generative adversarial framework for cold-start item recommendation}. In \bibinfo{booktitle}{\emph{Proceedings of the 45th International ACM SIGIR Conference on Research and Development in Information Retrieval}}. \bibinfo{pages}{2565--2571}.
\newblock
\href{https://doi.org/10.1145/3477495.3531897}{doi:\nolinkurl{10.1145/3477495.3531897}}


\bibitem[Gusak et~al\mbox{.}(2025)]%
        {seqsplits2025}
\bibfield{author}{\bibinfo{person}{Danil Gusak}, \bibinfo{person}{Anna Volodkevich}, \bibinfo{person}{Anton Klenitskiy}, \bibinfo{person}{Alexey Vasilev}, {and} \bibinfo{person}{Evgeny Frolov}.} \bibinfo{year}{2025}\natexlab{}.
\newblock \showarticletitle{Time to Split: Exploring Data Splitting Strategies for Offline Evaluation of Sequential Recommenders}. In \bibinfo{booktitle}{\emph{Proceedings of the 19th ACM Conference on Recommender Systems}}.
\newblock
\href{https://doi.org/10.1145/3705328.3748164}{doi:\nolinkurl{10.1145/3705328.3748164}}


\bibitem[Harte et~al\mbox{.}(2023)]%
        {harte2023leveraging}
\bibfield{author}{\bibinfo{person}{Jesse Harte}, \bibinfo{person}{Wouter Zorgdrager}, \bibinfo{person}{Panos Louridas}, \bibinfo{person}{Asterios Katsifodimos}, \bibinfo{person}{Dietmar Jannach}, {and} \bibinfo{person}{Marios Fragkoulis}.} \bibinfo{year}{2023}\natexlab{}.
\newblock \showarticletitle{Leveraging large language models for sequential recommendation}. In \bibinfo{booktitle}{\emph{Proceedings of the 17th ACM Conference on Recommender Systems}}. \bibinfo{pages}{1096--1102}.
\newblock
\href{https://doi.org/10.1145/3604915.3610639}{doi:\nolinkurl{10.1145/3604915.3610639}}


\bibitem[Hidasi and Czapp(2023)]%
        {hidasi2023widespread}
\bibfield{author}{\bibinfo{person}{Bal{\'a}zs Hidasi} {and} \bibinfo{person}{{\'A}d{\'a}m~Tibor Czapp}.} \bibinfo{year}{2023}\natexlab{}.
\newblock \showarticletitle{Widespread flaws in offline evaluation of recommender systems}. In \bibinfo{booktitle}{\emph{Proceedings of the 17th acm conference on recommender systems}}. \bibinfo{pages}{848--855}.
\newblock
\href{https://doi.org/10.1145/3604915.3608839}{doi:\nolinkurl{10.1145/3604915.3608839}}


\bibitem[Huang et~al\mbox{.}(2023)]%
        {huang2023aligning}
\bibfield{author}{\bibinfo{person}{Feiran Huang}, \bibinfo{person}{Zefan Wang}, \bibinfo{person}{Xiao Huang}, \bibinfo{person}{Yufeng Qian}, \bibinfo{person}{Zhetao Li}, {and} \bibinfo{person}{Hao Chen}.} \bibinfo{year}{2023}\natexlab{}.
\newblock \showarticletitle{Aligning distillation for cold-start item recommendation}. In \bibinfo{booktitle}{\emph{Proceedings of the 46th International ACM SIGIR Conference on Research and Development in Information Retrieval}}. \bibinfo{pages}{1147--1157}.
\newblock
\href{https://doi.org/10.1145/3539618.3591732}{doi:\nolinkurl{10.1145/3539618.3591732}}


\bibitem[Ji et~al\mbox{.}(2023)]%
        {ji2023critical}
\bibfield{author}{\bibinfo{person}{Yitong Ji}, \bibinfo{person}{Aixin Sun}, \bibinfo{person}{Jie Zhang}, {and} \bibinfo{person}{Chenliang Li}.} \bibinfo{year}{2023}\natexlab{}.
\newblock \showarticletitle{A critical study on data leakage in recommender system offline evaluation}.
\newblock \bibinfo{journal}{\emph{ACM Transactions on Information Systems}} \bibinfo{volume}{41}, \bibinfo{number}{3} (\bibinfo{year}{2023}), \bibinfo{pages}{1--27}.
\newblock
\href{https://doi.org/10.1145/3569930}{doi:\nolinkurl{10.1145/3569930}}


\bibitem[Jin et~al\mbox{.}(2023)]%
        {jin2023amazon}
\bibfield{author}{\bibinfo{person}{Wei Jin}, \bibinfo{person}{Haitao Mao}, \bibinfo{person}{Zheng Li}, \bibinfo{person}{Haoming Jiang}, \bibinfo{person}{Chen Luo}, \bibinfo{person}{Hongzhi Wen}, \bibinfo{person}{Haoyu Han}, \bibinfo{person}{Hanqing Lu}, \bibinfo{person}{Zhengyang Wang}, \bibinfo{person}{Ruirui Li}, {et~al\mbox{.}}} \bibinfo{year}{2023}\natexlab{}.
\newblock \showarticletitle{Amazon-m2: A multilingual multi-locale shopping session dataset for recommendation and text generation}.
\newblock \bibinfo{journal}{\emph{Advances in Neural Information Processing Systems}}  \bibinfo{volume}{36} (\bibinfo{year}{2023}), \bibinfo{pages}{8006--8026}.
\newblock
\urldef\tempurl%
\url{https://dl.acm.org/doi/10.5555/3666122.3666473}
\showURL{%
\tempurl}


\bibitem[Kang and McAuley(2018)]%
        {kang2018self}
\bibfield{author}{\bibinfo{person}{Wang-Cheng Kang} {and} \bibinfo{person}{Julian McAuley}.} \bibinfo{year}{2018}\natexlab{}.
\newblock \showarticletitle{Self-attentive sequential recommendation}. In \bibinfo{booktitle}{\emph{2018 IEEE international conference on data mining (ICDM)}}. IEEE, \bibinfo{pages}{197--206}.
\newblock
\href{https://doi.org/10.1109/ICDM.2018.00035}{doi:\nolinkurl{10.1109/ICDM.2018.00035}}


\bibitem[Klenitskiy and Vasilev(2023)]%
        {klenitskiy2023turning}
\bibfield{author}{\bibinfo{person}{Anton Klenitskiy} {and} \bibinfo{person}{Alexey Vasilev}.} \bibinfo{year}{2023}\natexlab{}.
\newblock \showarticletitle{Turning dross into gold loss: is bert4rec really better than sasrec?}. In \bibinfo{booktitle}{\emph{Proceedings of the 17th ACM Conference on Recommender Systems}}. \bibinfo{pages}{1120--1125}.
\newblock
\href{https://doi.org/10.1145/3604915.3610644}{doi:\nolinkurl{10.1145/3604915.3610644}}


\bibitem[Klenitskiy et~al\mbox{.}(2024)]%
        {klenitskiy2024does}
\bibfield{author}{\bibinfo{person}{Anton Klenitskiy}, \bibinfo{person}{Anna Volodkevich}, \bibinfo{person}{Anton Pembek}, {and} \bibinfo{person}{Alexey Vasilev}.} \bibinfo{year}{2024}\natexlab{}.
\newblock \showarticletitle{Does it look sequential? an analysis of datasets for evaluation of sequential recommendations}. In \bibinfo{booktitle}{\emph{Proceedings of the 18th ACM Conference on Recommender Systems}}. \bibinfo{pages}{1067--1072}.
\newblock
\href{https://doi.org/10.1145/3640457.3688195}{doi:\nolinkurl{10.1145/3640457.3688195}}


\bibitem[Li et~al\mbox{.}(2023)]%
        {li2023text}
\bibfield{author}{\bibinfo{person}{Jiacheng Li}, \bibinfo{person}{Ming Wang}, \bibinfo{person}{Jin Li}, \bibinfo{person}{Jinmiao Fu}, \bibinfo{person}{Xin Shen}, \bibinfo{person}{Jingbo Shang}, {and} \bibinfo{person}{Julian McAuley}.} \bibinfo{year}{2023}\natexlab{}.
\newblock \showarticletitle{Text is all you need: Learning language representations for sequential recommendation}. In \bibinfo{booktitle}{\emph{Proceedings of the 29th ACM SIGKDD Conference on Knowledge Discovery and Data Mining}}. \bibinfo{pages}{1258--1267}.
\newblock
\href{https://doi.org/10.1145/3580305.3599519}{doi:\nolinkurl{10.1145/3580305.3599519}}


\bibitem[McAuley et~al\mbox{.}(2015)]%
        {mcauley2015image}
\bibfield{author}{\bibinfo{person}{Julian McAuley}, \bibinfo{person}{Christopher Targett}, \bibinfo{person}{Qinfeng Shi}, {and} \bibinfo{person}{Anton Van Den~Hengel}.} \bibinfo{year}{2015}\natexlab{}.
\newblock \showarticletitle{Image-based recommendations on styles and substitutes}. In \bibinfo{booktitle}{\emph{Proceedings of the 38th international ACM SIGIR conference on research and development in information retrieval}}. \bibinfo{pages}{43--52}.
\newblock
\href{https://doi.org/10.1145/2766462.2767755}{doi:\nolinkurl{10.1145/2766462.2767755}}


\bibitem[Shalaby et~al\mbox{.}(2022)]%
        {shalaby2022m2trec}
\bibfield{author}{\bibinfo{person}{Walid Shalaby}, \bibinfo{person}{Sejoon Oh}, \bibinfo{person}{Amir Afsharinejad}, \bibinfo{person}{Srijan Kumar}, {and} \bibinfo{person}{Xiquan Cui}.} \bibinfo{year}{2022}\natexlab{}.
\newblock \showarticletitle{M2TRec: Metadata-aware Multi-task Transformer for Large-scale and Cold-start free Session-based Recommendations}. In \bibinfo{booktitle}{\emph{Proceedings of the 16th ACM Conference on Recommender Systems}}. \bibinfo{pages}{573--578}.
\newblock
\href{https://doi.org/10.1145/3523227.3551477}{doi:\nolinkurl{10.1145/3523227.3551477}}


\bibitem[Shevchenko et~al\mbox{.}(2024)]%
        {shevchenko2024variability}
\bibfield{author}{\bibinfo{person}{Valeriy Shevchenko}, \bibinfo{person}{Nikita Belousov}, \bibinfo{person}{Alexey Vasilev}, \bibinfo{person}{Vladimir Zholobov}, \bibinfo{person}{Artyom Sosedka}, \bibinfo{person}{Natalia Semenova}, \bibinfo{person}{Anna Volodkevich}, \bibinfo{person}{Andrey Savchenko}, {and} \bibinfo{person}{Alexey Zaytsev}.} \bibinfo{year}{2024}\natexlab{}.
\newblock \showarticletitle{From Variability to Stability: Advancing RecSys Benchmarking Practices}. In \bibinfo{booktitle}{\emph{Proceedings of the 30th ACM SIGKDD Conference on Knowledge Discovery and Data Mining}} (Barcelona, Spain) \emph{(\bibinfo{series}{KDD '24})}. \bibinfo{publisher}{Association for Computing Machinery}, \bibinfo{address}{New York, NY, USA}, \bibinfo{pages}{5701–5712}.
\newblock
\showISBNx{9798400704901}
\href{https://doi.org/10.1145/3637528.3671655}{doi:\nolinkurl{10.1145/3637528.3671655}}


\bibitem[Sun(2023)]%
        {sun2023take}
\bibfield{author}{\bibinfo{person}{Aixin Sun}.} \bibinfo{year}{2023}\natexlab{}.
\newblock \showarticletitle{Take a fresh look at recommender systems from an evaluation standpoint}. In \bibinfo{booktitle}{\emph{Proceedings of the 46th International ACM SIGIR Conference on Research and Development in Information Retrieval}}. \bibinfo{pages}{2629--2638}.
\newblock
\href{https://doi.org/10.1145/3539618.3591931}{doi:\nolinkurl{10.1145/3539618.3591931}}


\bibitem[Sun et~al\mbox{.}(2019)]%
        {sun2019bert4rec}
\bibfield{author}{\bibinfo{person}{Fei Sun}, \bibinfo{person}{Jun Liu}, \bibinfo{person}{Jian Wu}, \bibinfo{person}{Changhua Pei}, \bibinfo{person}{Xiao Lin}, \bibinfo{person}{Wenwu Ou}, {and} \bibinfo{person}{Peng Jiang}.} \bibinfo{year}{2019}\natexlab{}.
\newblock \showarticletitle{BERT4Rec: Sequential recommendation with bidirectional encoder representations from transformer}. In \bibinfo{booktitle}{\emph{Proceedings of the 28th ACM international conference on information and knowledge management}}. \bibinfo{pages}{1441--1450}.
\newblock
\href{https://doi.org/10.1145/3357384.3357895}{doi:\nolinkurl{10.1145/3357384.3357895}}


\bibitem[Sun et~al\mbox{.}(2020)]%
        {sun2020we}
\bibfield{author}{\bibinfo{person}{Zhu Sun}, \bibinfo{person}{Di Yu}, \bibinfo{person}{Hui Fang}, \bibinfo{person}{Jie Yang}, \bibinfo{person}{Xinghua Qu}, \bibinfo{person}{Jie Zhang}, {and} \bibinfo{person}{Cong Geng}.} \bibinfo{year}{2020}\natexlab{}.
\newblock \showarticletitle{Are we evaluating rigorously? benchmarking recommendation for reproducible evaluation and fair comparison}. In \bibinfo{booktitle}{\emph{Proceedings of the 14th ACM Conference on Recommender Systems}}. \bibinfo{pages}{23--32}.
\newblock
\href{https://doi.org/10.1145/3383313.3412489}{doi:\nolinkurl{10.1145/3383313.3412489}}


\bibitem[Tamm and Aljanaki(2024)]%
        {tamm2024comparative}
\bibfield{author}{\bibinfo{person}{Yan-Martin Tamm} {and} \bibinfo{person}{Anna Aljanaki}.} \bibinfo{year}{2024}\natexlab{}.
\newblock \showarticletitle{Comparative Analysis of Pretrained Audio Representations in Music Recommender Systems}. In \bibinfo{booktitle}{\emph{Proceedings of the 18th ACM Conference on Recommender Systems}}. \bibinfo{pages}{934--938}.
\newblock
\href{https://doi.org/10.1145/3640457.3688172}{doi:\nolinkurl{10.1145/3640457.3688172}}


\bibitem[Van~den Oord et~al\mbox{.}(2013)]%
        {van2013deep}
\bibfield{author}{\bibinfo{person}{Aaron Van~den Oord}, \bibinfo{person}{Sander Dieleman}, {and} \bibinfo{person}{Benjamin Schrauwen}.} \bibinfo{year}{2013}\natexlab{}.
\newblock \showarticletitle{Deep content-based music recommendation}.
\newblock \bibinfo{journal}{\emph{Advances in neural information processing systems}}  \bibinfo{volume}{26} (\bibinfo{year}{2013}).
\newblock
\urldef\tempurl%
\url{https://dl.acm.org/doi/10.5555/2999792.2999907}
\showURL{%
\tempurl}


\bibitem[Vasilev et~al\mbox{.}(2024)]%
        {vasilev2024replay}
\bibfield{author}{\bibinfo{person}{Alexey Vasilev}, \bibinfo{person}{Anna Volodkevich}, \bibinfo{person}{Denis Kulandin}, \bibinfo{person}{Tatiana Bysheva}, {and} \bibinfo{person}{Anton Klenitskiy}.} \bibinfo{year}{2024}\natexlab{}.
\newblock \showarticletitle{RePlay: a Recommendation Framework for Experimentation and Production Use}. In \bibinfo{booktitle}{\emph{Proceedings of the 18th ACM Conference on Recommender Systems}}. \bibinfo{pages}{1191--1194}.
\newblock
\href{https://doi.org/10.1145/3640457.3691701}{doi:\nolinkurl{10.1145/3640457.3691701}}


\bibitem[Volkovs et~al\mbox{.}(2017)]%
        {volkovs2017dropoutnet}
\bibfield{author}{\bibinfo{person}{Maksims Volkovs}, \bibinfo{person}{Guangwei Yu}, {and} \bibinfo{person}{Tomi Poutanen}.} \bibinfo{year}{2017}\natexlab{}.
\newblock \showarticletitle{Dropoutnet: Addressing cold start in recommender systems}.
\newblock \bibinfo{journal}{\emph{Advances in neural information processing systems}}  \bibinfo{volume}{30} (\bibinfo{year}{2017}).
\newblock
\urldef\tempurl%
\url{https://dl.acm.org/doi/10.5555/3295222.3295249}
\showURL{%
\tempurl}


\bibitem[Wang et~al\mbox{.}(2022)]%
        {wang2022text}
\bibfield{author}{\bibinfo{person}{Liang Wang}, \bibinfo{person}{Nan Yang}, \bibinfo{person}{Xiaolong Huang}, \bibinfo{person}{Binxing Jiao}, \bibinfo{person}{Linjun Yang}, \bibinfo{person}{Daxin Jiang}, \bibinfo{person}{Rangan Majumder}, {and} \bibinfo{person}{Furu Wei}.} \bibinfo{year}{2022}\natexlab{}.
\newblock \showarticletitle{Text Embeddings by Weakly-Supervised Contrastive Pre-training}.
\newblock \bibinfo{journal}{\emph{arXiv preprint arXiv:2212.03533}} (\bibinfo{year}{2022}).
\newblock
\href{https://doi.org/10.48550/arXiv.2212.03533}{doi:\nolinkurl{10.48550/arXiv.2212.03533}}


\bibitem[Wang et~al\mbox{.}(2024)]%
        {wang2024language}
\bibfield{author}{\bibinfo{person}{Shiyu Wang}, \bibinfo{person}{Hao Ding}, \bibinfo{person}{Yupeng Gu}, \bibinfo{person}{Sergul Aydore}, \bibinfo{person}{Kousha Kalantari}, {and} \bibinfo{person}{Branislav Kveton}.} \bibinfo{year}{2024}\natexlab{}.
\newblock \showarticletitle{Language-Model Prior Overcomes Cold-Start Items}.
\newblock \bibinfo{journal}{\emph{arXiv preprint arXiv:2411.09065}} (\bibinfo{year}{2024}).
\newblock
\href{https://doi.org/10.48550/arXiv.2411.09065}{doi:\nolinkurl{10.48550/arXiv.2411.09065}}


\bibitem[Wei et~al\mbox{.}(2021)]%
        {wei2021contrastive}
\bibfield{author}{\bibinfo{person}{Yinwei Wei}, \bibinfo{person}{Xiang Wang}, \bibinfo{person}{Qi Li}, \bibinfo{person}{Liqiang Nie}, \bibinfo{person}{Yan Li}, \bibinfo{person}{Xuanping Li}, {and} \bibinfo{person}{Tat-Seng Chua}.} \bibinfo{year}{2021}\natexlab{}.
\newblock \showarticletitle{Contrastive learning for cold-start recommendation}. In \bibinfo{booktitle}{\emph{Proceedings of the 29th ACM international conference on multimedia}}. \bibinfo{pages}{5382--5390}.
\newblock
\href{https://doi.org/10.1145/3474085.3475665}{doi:\nolinkurl{10.1145/3474085.3475665}}


\bibitem[Zhu et~al\mbox{.}(2021)]%
        {zhu2021learning}
\bibfield{author}{\bibinfo{person}{Yongchun Zhu}, \bibinfo{person}{Ruobing Xie}, \bibinfo{person}{Fuzhen Zhuang}, \bibinfo{person}{Kaikai Ge}, \bibinfo{person}{Ying Sun}, \bibinfo{person}{Xu Zhang}, \bibinfo{person}{Leyu Lin}, {and} \bibinfo{person}{Juan Cao}.} \bibinfo{year}{2021}\natexlab{}.
\newblock \showarticletitle{Learning to warm up cold item embeddings for cold-start recommendation with meta scaling and shifting networks}. In \bibinfo{booktitle}{\emph{Proceedings of the 44th International ACM SIGIR Conference on Research and Development in Information Retrieval}}. \bibinfo{pages}{1167--1176}.
\newblock
\href{https://doi.org/10.1145/3404835.3462843}{doi:\nolinkurl{10.1145/3404835.3462843}}


\end{thebibliography}

\end{document}